\documentclass{aastex61}
\usepackage{graphicx}
\usepackage{amsmath,bm}

\shorttitle{Eruption of hot coronal channel}
\shortauthors{Mitra et al.}

\begin{document}

\title{Pre-flare processes, flux rope activation, large-scale eruption and associated X-class flare from the active region NOAA 11875}

\correspondingauthor{Prabir K. Mitra}
\email{prabir@prl.res.in}

\author{Prabir K. Mitra}
\affiliation{Udaipur Solar Observatory, Physical Research Laboratory, Udaipur 313 001, India}
\affiliation{Department of Physics, Gujarat University, Ahmedabad 380 009, India}

\author{Bhuwan Joshi}
\affiliation{Udaipur Solar Observatory, Physical Research Laboratory, Udaipur 313 001, India}

\begin{abstract}
We present a multi-wavelength analysis of the eruption of a hot coronal channel associated with an X1.0 flare (SOL2013-10-28T02:03) from the active region NOAA 11875 by combining observations from AIA/\textit{SDO}, HMI/\textit{SDO}, \textit{RHESSI}, and HiRAS. EUV images at high coronal temperatures indicated the presence of a hot channel at the core of the active region from the early pre-flare phase evidencing the pre-existence of a quasi-stable magnetic flux rope. The hot channel underwent an activation phase after a localized and prolonged pre-flare event occurring adjacent to one of its footpoints. Subsequently, the flux rope continued to rise slowly for $\approx$16 min during which soft X-ray flux gradually built-up characterizing a distinct precursor phase. The flux rope transitioned from the state of slow rise to the eruptive motion with the onset of the impulsive phase of the X1.0 flare. The eruptive expansion of the hot channel is accompanied by a series of type III radio bursts in association with impulsive rise of strong hard X-ray non-thermal emissions that included explicit hard X-ray sources of energies up to $\approx$50 keV from the coronal loops and $\approx$100 keV from their footpoint locations. Our study contains evidence that pre-flare activity occurring within the spatial extent of a stable flux rope can destabilize it toward eruption. Moreover, sudden transition of the flux rope from the state of slow rise to fast acceleration precisely bifurcated the precursor and the impulsive phases of the flare which points toward a feedback relationship between early CME dynamics and the strength of the large-scale magnetic reconnection.
\end{abstract}

\keywords{Sun: activity --- Sun: corona --- Sun: filaments, prominences --- Sun: flares --- Sun: X-rays, gamma rays}

\section{Introduction} \label{intro} 
Magnetic flux ropes (MFRs) are defined as sets of magnetic field lines which are twisted along a common central axis usually more than once \citep{Gibson2006}. MFRs are observed to be levitated along the polarity inversion line (PIL) with the two legs tied in the opposite polarity regions. With a high storage of magnetic free energy, MFRs are one of the essential ingredients of large-scale solar transient phenomena i.e., flares and coronal mass ejections (CMEs) \citep{Jing2010}. Earth-directed CMEs cause geomagnetic storms which may disrupt the spacecrafts stationed in near-Earth orbits and the communication system on Earth \citep{Forsyth2006, Webb2012}. As the core of CMEs, MFRs have been studied extensively in the recent years and remain one of the most contemporary topics in solar physics.

Different solar features viz., filaments, prominences, filament channels, hot coronal channels, etc. essentially form the observational counterparts of MFRs. Filaments are thread like structures found as dark features in the chromospheric images of the Sun. These are cool, dense material supported in the corona against gravity by sheared magnetic field with a dip \citep{Antiochos1994} or helically twisted field lines \citep{Priest1989}. When these structures are observed on the limb, they are called prominences since they appear bright compared to the background. Filament channels are long-lived, narrow lanes between extended areas of magnetic field of opposite polarities where a filament (or, prominence) can be formed \citep{Gaizauskas1997, Engvold1997}. Hot coronal channels were first reported by \citet{Zhang2012, Cheng2013} as coherent structures found in high temperature pass-band filter images. Hot channels are often found to appear in association with coronal sigmoids that are observed in EUV images \citep{Cheng2014b, Joshi2017, Joshi2018, Mitra2018}. It is noteworthy that one footpoint of the hot coronal channel may originate in a strong magnetic field region with the other footpoint terminating in a weaker magnetic field \citep{Cheng2016}. Below the hot channel, a filament channel is often observed confirming its association with magnetic flux ropes \citep{Chenb2014}. An observational survey performed by \citet{Nindos2015} revealed that almost half of the major eruptive flares are associated with a pre-flare hot channel.

Successful eruption of an MFR is essential for the origin of a CME. According to the `standard flare model' \citep[also known as CSHKP model;][]{Carmichael1964, Sturrock1966, Hirayama1974, Kopp1976}, once an MFR undergoes eruptive expansion, magnetic reconnection sets in between the inflow magnetic field lines. During magnetic reconnection, the stored magnetic energy is released in the form of intense heating within a localized region and non-thermal emissions \citep{Shibata2011}. Multi-wavelength observations of solar flares demonstrate complex temporal, spatial, and spectral variabilities in the energy release processes \citep{Fletcher2011, Benz2017}. CSHKP model is successful in explaining most the commonly observed signatures of a flare, such as, flare ribbons; looptop and footpoint X-ray sources; cusp structure following the passage of the filament; post-flare loop arcade; etc. However, it remains silent to the topics of flux rope formation and triggering of the eruption process. It is also noteworthy that in view of the spatial evolution of the looptop and footpoint sources during the early impulsive phases, many flares do not comply with the CSHKP model \citep[see e.g.,][]{Veronig2006, Joshi2009, Joshi2012}. Further, the morphological and dynamical evolution of flare ribbons could be very complex \citep[see e.g.,][]{Sui2003, Dalmasse2015, Joshi2017, Mitra2018, Joshi2019}.

Triggering mechanism of flux ropes is one of the most debated topics among the solar physicists. Years of observation and corresponding theoretical understanding have led to two different classes of models for triggering mechanism: reconnection based triggering and ideal instability. Two representative reconnection based triggering mechanisms-- tether cutting and breakout models-- rely on different pre-flare magnetic configurations. The tether cutting model involves two highly sheared arcades in a bipolar active region where triggering occurs in the form of initial reconnection beneath the sheared arcades when a newly emerged bipole interacts with them \citep{Moore1992, Moore2001}. The breakout model, on the other hand, requires a complex multipolar active region with one or more pre-existing magnetic nulls well above the core field containing a sheared arcade. Initial reconnection, according to the breakout model, occurs in those magnetic nulls which reduces downward tension on the flux rope setting it in an eruptive motion \citep{Antiochos1999}. Over the years, several case studies have provided support for the tether cutting model \citep[see e.g.,][]{Liu2007, Liu2013, Chen2014, Xue2017, Chen2018, Yang2018}. Similarly, several observational studied and simulations have supported the breakout model of solar eruption \citep[see e.g.,][]{Manoharan2003, Gary2004, Joshi2007, Aurass2011, ChenY2016, Mitra2018}.

The ideal instability models for triggering mechanism do not rely on initial reconnection for eruption of MFRs. Two basic instability models-- torus instability \citep{Torok2006} and kink instability \citep{Torok2004}-- predicts the onset of eruption of an MFR when some critical values are reached. According to the torus instability model, an MFR may attain eruptive expansion if the overlying magnetic field experiences a sharp decay with height. The critical value is decided by the parameter `decay index' ($n=-\frac{log(B_{ex})}{log(z)}$, where $B_{ex}$ and z are the overlying magnetic field and height, respectively) and several studies have shown the critical value of decay index to lie within the range [1.1--1.5] \citep{Demoulin2010, Olmedo2010}. The kink instability model suggests eruption of an MFR if its twist increases beyond a critical value of $\approx$3.5$\pi$. However, several studies have shown that both these two instability criteria may lead to failed eruption of an MFR \citep{Liu2008, Song2014} and often both the torus and kink instabilities are simultaneously required for a successful eruption of an MFR \citep{Liu2008, Vemareddy2014}.

During a flare, huge amount of magnetic energy is released in the form of thermal and non-thermal energies radiating across the entire electromagnetic spectrum i.e., from $\gamma$-rays to radio waves. Thermal signatures of flares include optical, ultra-violet (UV), extreme ultra-violet (EUV), and soft X-ray (SXR) brightnenings from post-reconnection loop arcades and flare ribbons \citep[see review by,][]{Fletcher2011}. Non-thermal signatures of flares include hard X-ray (HXR), microwave, and radio bursts which individually carries evidences of different physical mechanisms during a flare \citep{Krucker2008, White2011}. Careful observation and rightful analysis of flare HXR sources can be used to constraint the location of magnetic reconnection and particle acceleration \citep{Masuda1994, Benz2017} whereas type III and type II radio bursts can indirectly provide information on the restructuring of coronal field and early phases of CME propagation \citep{Wild1950, Cairns2003, Reid2014}.

In this work, we present a multi-wavelength study of the eruption of a hot coronal channel which led to an X1.0 flare, using high cadence and high spatial resolution observations from the Atmospheric Imaging Assembly \citep[AIA;][]{Lemen2012}. The flaring region was associated with multiple hard X-ray sources of energies up to $\approx$100 keV. The Hiraiso Radio Spectrograph \citep[HiRAS;][]{Kondo1995} revealed complex set of metric radio bursts: a series of type III during the early rise phase, a set of prominent split-band type II harmonic band around the peak phase and a type IV burst during the decay phase.
An important objective of the study lies in understanding the pre-flare processes and their relevance to the subsequent phases of CME initiation and main flaring event. The study provides an unambiguous detection of a pre-existing MFR in the active region and shows its various activation stages that lead to a standard CME-producing X-class flares. The availability of solar X-ray observations from the \textit{Reuven Ramaty High Energy Solar Spectroscopic Imager} \citep[\textit{RHESSI};][]{Lin2002} during part of the pre-flare phase and most of the X-class flare provided us with an opportunity to explore the sites of energization, heating, and particle acceleration. Section \ref{data} provides details of the observational data and analysis techniques used in this article. Multi-wavelength results involving EUV, X-ray and radio observations are explained in Section \ref{mulob}. We discuss and interpret our results in Section \ref{disc}. The major highlights and conclusions of the study are provided in Section \ref{summary}.

\section{Observational Data and Methods} \label{data}
For observing the Sun in (E)UV wavelengths, we use high resolution (0$\farcs$6 pixel$^{-1}$), 4096$\times$4096 pixel full disk observations from the Atmospheric Imaging Assembly \citep[AIA;][]{Lemen2012} on board the \textit{Solar Dynamics Observatory} \citep[\textit{SDO};][]{Pesnell2012}. AIA observes the Sun in 7 EUV channels (94 \AA , 131 \AA , 171 \AA , 193 \AA , 211 \AA , 304 \AA , and 335 \AA ), 2 UV channels (1600 \AA\ and 1700 \AA) and one white light channel (4500 \AA ). Temporal cadences are 12 s for the EUV filters, 24 s for the UV filters and 3600 s for the white light filter. For photospheric observation, we use 45 s cadence line of sight (LOS) magnetograms and intensity images observed by Helioseismic and Magnetic Imager \citep[HMI;][]{Schou2012} on board \textit{SDO}, which takes continuous full disk observation of the solar photosphere with spatial resolution of 0$\farcs$5 pixel$^{-1}$.

Coronal Mass Ejection (CME) associated with the erupting hot channel was observed by the C2 and C3 instruments of the Large Angle and Spectrometric Coronagraph \citep[LASCO;][]{Brueckner1995} on board the \textit{Solar and Heliospheric Observatory} \citep[\textit{SOHO};][]{Domingo1995}. C2 and C3 are white light coronagraphs imaging from 1.5 to 6 $R_\odot$ and from 3.7 to 30 $R_\odot$, respectively.

The \textit{Reuven Ramaty High Energy Solar Spectroscopic Imager} \citep[\textit{RHESSI};][]{Lin2002} provided information about the X-ray sources associated with the flaring region. \textit{RHESSI} observes the full Sun with an unprecedented spatial resolution (as fine as $\sim$2$\farcs3$) and energy resolution (1--5 keV) in the energy range 3 keV--17 MeV. For imaging of the X-ray sources using \textit{RHESSI} observation, we use the PIXON algorithm \citep{Metcalf1996} with the natural weighting scheme for front detector segments 2--9 (excluding 7). For construction of \textit{RHESSI} spectra, we use the front detector segments 1--9 excluding 2 and 7 \citep[which have lower energy resolution and high threshold energies, respectively; see][]{Smith2002, Holman2011}. The spectra were deconvolved with the full detector response matrix \citep[i.e., off-diagonal elements were included;][]{Smith2002}.

Dynamic radio spectra within the frequency range 50--500 MHz was obtained from the Hiraiso Radio Spectrograph \citep[HiRAS;][]{Kondo1995}, which is operated by the National Institute of Information and Communications Technology, Japan\footnote{\url{http://sunbase.nict.go.jp/solar/denpa/index.html}}.

\section{Multi-wavelength Observations and Results} \label{mulob}

\subsection{Active region NOAA 11875} \label{S_intro}
The active region (AR) NOAA 11875 emerged on the eastern limb of the Sun on 2013 October 16 as a $\beta\gamma$-type sunspot\footnote{\url{www.helioviewer.org}}. On October 22, it transformed into a more complex $\beta\gamma\delta$-type and remained so until its disappearance from the western limb of the Sun on October 30. In its lifetime, it produced 2 X-class, 11 M-class besides many C-class flares\footnote{\url{http://www.lmsal.com/solarsoft/latest_events_archive.html}}.

\begin{figure}
   \centering
   \epsscale{1.1}
   \plotone{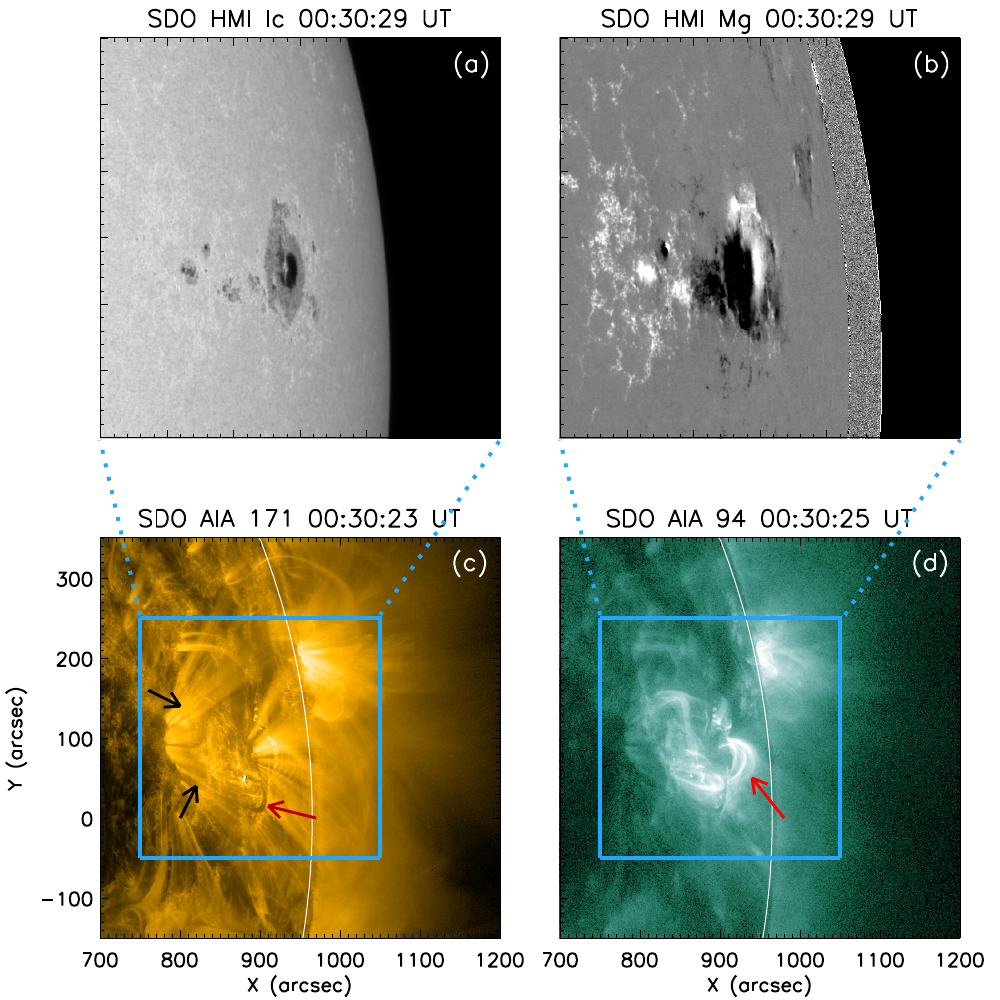}
\caption{Panel (a): HMI white light image of the active region NOAA 11875 prior to the eruptive flare. Panel (b): Co-temporal HMI LOS magnetogram of AR 11875. Panels (c)--(d): AIA EUV images of the AR in 171 \AA\ and 94 \AA\ , respectively. FOV of panels (a) and (b) is indicated by the sky colored boxes in panels (c) and (d). The black arrows in panel (c) indicate few open coronal lines. The brown arrow indicate the filament during the pre-flare phase. The red arrow in panel (d) indicate the hot coronal loops near the filament.}
\label{I_intro}
\end{figure}

We have studied the active region NOAA 11875 on 2013 October 28 from 00:30 UT to 04:30 UT while it was situated close to the western limb of the Sun at the heliographic co-ordinates $\sim$N07W66\footnote{\url{https://www.solarmonitor.org/index.php?date=20131028&region=11875}}. In this duration, the active region underwent activation and eruption of a hot channel in association with an X1.0 flare. In Figure \ref{I_intro}, we present multi-wavelength view of the AR prior to the erupting event. The photospheric HMI white light image suggests that the active region was consisted of a major leading sunspot and few trailing spots of much smaller sizes (Figure \ref{I_intro}(a)). Comparison of a co-temporal HMI LOS magnetogram (Figure \ref{I_intro}(b)) with the while light image indicates that the major leading sunspot had a complex fine structure with multiple umbrae of opposite polarity within a common penumbra i.e., $\delta$-spots while the trailing spots were mostly of positive polarity. AIA 171 \AA\ image gives information about the coronal connectivities associated with the AR. Interestingly, we find a filament structure in the AR (indicated by the brown arrow in Figure \ref{I_intro}(c)) and a set of open field lines originating from the trailing part of the AR (marked by the black arrows in Figure \ref{I_intro}(c)). AIA 94 \AA\ image indicates a hot coronal channel near the filament (indicated by the red arrow in Figure \ref{I_intro}(d)) which forms the core of the AR. Notably, comparison of AIA 171 and 94 \AA\ images also indicate that the region of open field lines was also connected with the core of the AR (containing the filament) by a set of hot coronal loops.

\subsection{Evolutionary phases of the X1.0 flare}
\begin{figure}
   \centering
   \epsscale{1.1}
   \plotone{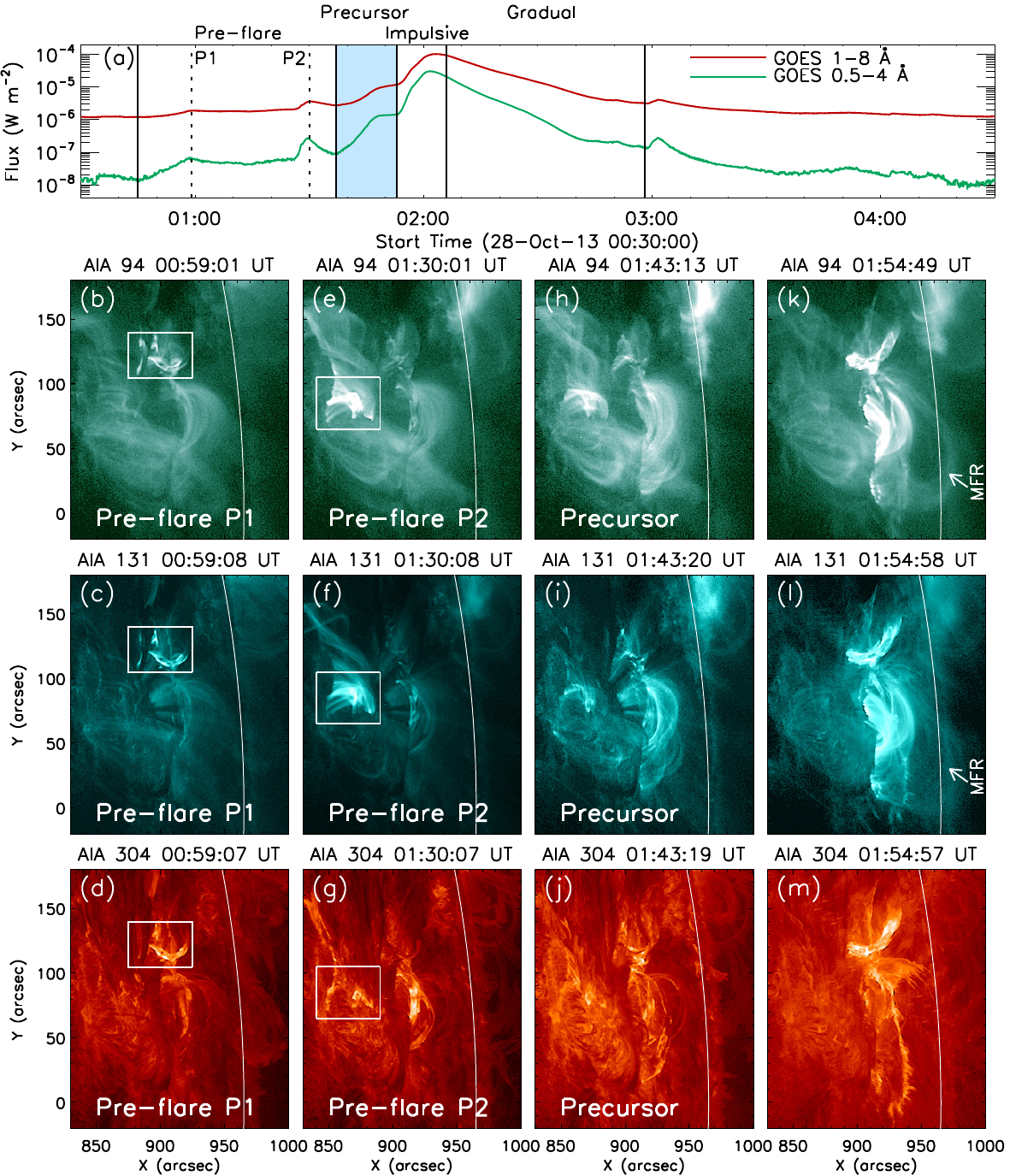}
\caption{Panel (a): GOES SXR flux variation in 1--8 \AA\ (red curve) and 0.5--4 \AA\ (green curve) channels on 2013 October 28 from 00:30 UT to 04:30 UT that includes different phases prior to and during the X1.0 flare. The pre-flare phase was characterized by two episodes of SXR flux enhancements with peaks `P1' and `P2' which are indicated by the dashed lines. The sky colored shaded area indicates the precursor phase when the activation of the hot channel (i.e., flux rope) began and it underwent a slow-rise phase. Panels (b)--(m): AIA EUV images of the active region NOAA 11875 in 94 \AA\ (panels (b), (e), (h), (k)), 131 \AA\ (panels (c), (f), (i), (l)), and 304 \AA\ (panels (d), (g), (j), (m)). These representative images show various phases of the X1.0 flare: two pre-flare peaks P1 and P2 (panels (b)--(d) and (e)--(g), respectively); precursor phase (panels (h)--(j)), and the eruption of the hot channel during the impulsive phase of the X1.0 flare (panels (k)--(m)). The white boxes in panels (b)--(d) indicate pre-flare activity during P1 from a complex localized region adjacent to the northern leg of the hot channel. The white boxes in panels (e)--(g) indicate pre-flare activity during P2 from the trailing part of the AR. The white arrows in panels (k) and (l) indicate the erupting flux rope.\\
(An animation of this figure is available.)}
\label{goes_aia}
\end{figure}

Figure \ref{goes_aia}(a) displays the time variation of the GOES SXR flux in 1--8 \AA\ and 0.5--4 \AA\ channels within our studied interval. The temporal evolution of the X1.0 flare can readily be summarized by three phases: a pre-flare phase showing two distinct episodes of SXR flux enhancement with peaks `P1' and `P2' (indicated by the two dotted lines in Figure \ref{goes_aia}(a)), a precursor phase and the X1.0 flare.
According to GOES, the onset of the impulsive phase of the X1.0 flare occurred at $\approx$01:53 UT which was characterized by rapid enhancement of flux in both the GOES X-ray channels and eruption of hot coronal channel. The flare reached at its peak flux at $\approx$02:02 UT which was followed by a gradual phase when SXR flux in both GOES channels decreased slowly.

During the first pre-flare event (P1), a complex structure, situated near the northern leg of the hot coronal channel (shown inside the boxes in Figures \ref{goes_aia}(b)--(d)) brightened up which could be observed in both high and low temperature AIA passband filter images. In Figures \ref{goes_aia}(b)--(d), we have plotted AIA 94 \AA , 131 \AA , and 304 \AA\ images, respectively, during the peak of the pre-flare event P1. During the second pre-flare event (P2), localized intensified brightness was observed from the trailing part of the AR (shown inside the white boxes in Figures \ref{goes_aia}(e)--(g)). AIA 94 \AA\ (Figure \ref{goes_aia}(e)) and AIA 131 \AA\ (Figure \ref{goes_aia}(f)) images during the pre-flare event P2 suggest that energy release involving a compact loop system probably led to the enhancement in GOES SXR flux during the second pre-flare event. From the co-temporal AIA 304 \AA\ image (Figure \ref{goes_aia}(g)), we observed two ribbon like brightening which outlined the footpoints of the intensified loops. In Figures \ref{goes_aia}(h)--(j), we have plotted AIA 94 \AA , 131 \AA , and 304 \AA\ images, respectively, during the precursor phase of the flare when the activation of the hot channel began and it underwent a slow rise phase. The subsequent eruption of the hot channel was clearly identified in AIA 94 \AA\ and 131 \AA\ images (indicated by white arrows in Figures \ref{goes_aia}(k) and (l)).

\begin{deluxetable*}{p{0.7cm}p{1.5cm}p{4cm}p{7.8cm}}
\tablenum{1}
\tablecaption{Summary of the evolutionary phases of the eruptive X1.0 flare}
\label{table}
\tablehead{
\colhead{Sr. No.} & \colhead{Phase} & \colhead{Interval} & \colhead{Remarks}
}

\startdata
\ \ \ 1 & \ \ Pre-flare & \ \ \ \ \ 00:45 UT-- 01:37 UT & Two peaks were observed in the GOES SXR flux at $\approx$00:59 UT (`P1') and $\approx$01:30 UT (`P2'). \\
\ \ \ 2 & \ Precursor & \ \ \ \ \ 01:37 UT-- 01:53 UT & Activation of the hot coronal channel began and it underwent a slow rise. \\
\ \ \ 3 & \ Impulsive & \ \ \ \ \ 01:53 UT-- 02:02 UT & The eruption of the hot channel experienced rapid acceleration; multiple X-ray sources were observed from footpoint and looptop locations.\\
\ \ \ 4 & \ \ Gradual & \ \ \ \ \ 02:02 UT-- 02:58 UT & Dense post-flare arcade was formed.\\
\enddata
\end{deluxetable*}

\begin{figure}
   \centering
   \epsscale{1.1}
   \plotone{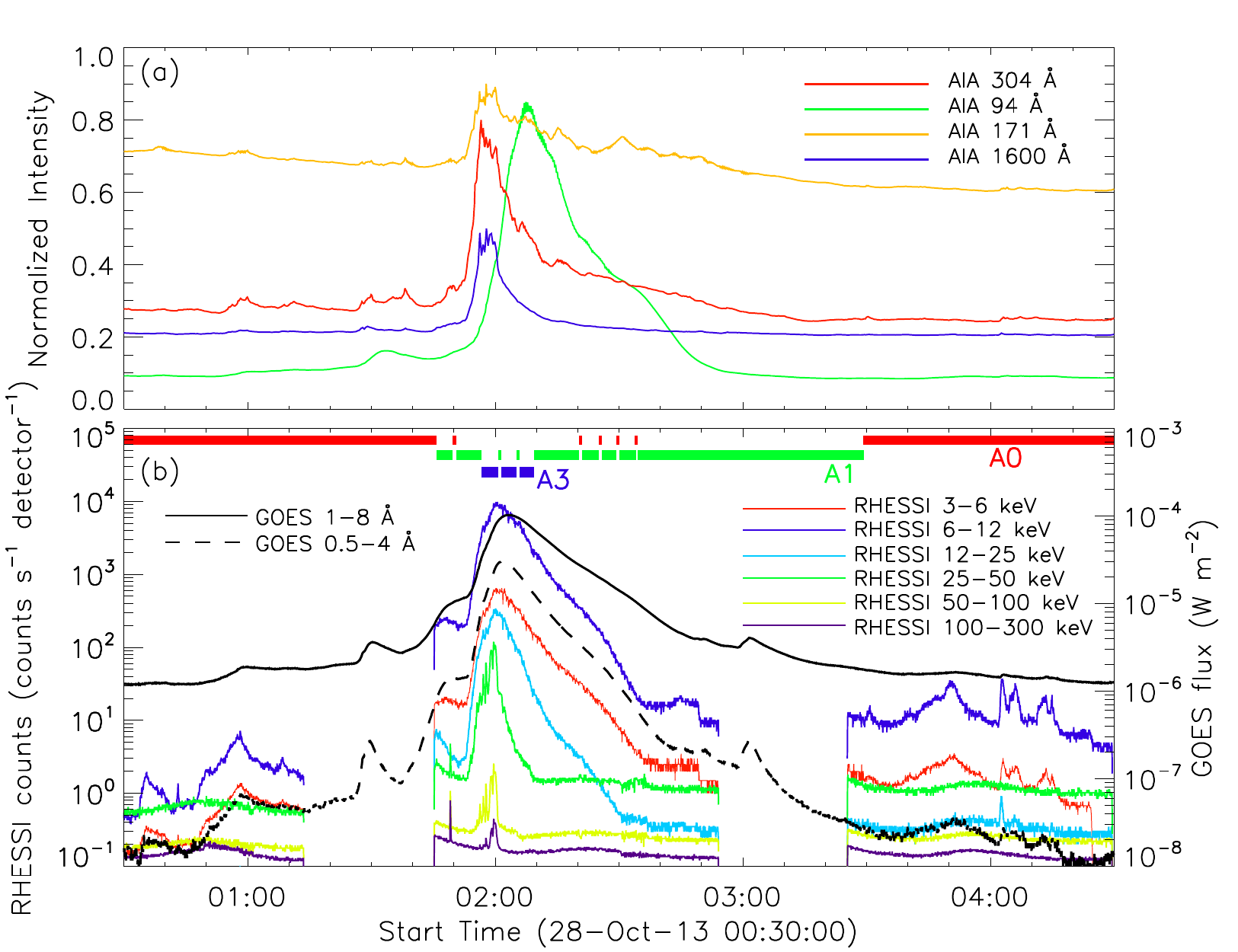}
\caption{Panel (a): AIA lightcurves on 2013 October 28 during 00:30--04:30 UT normalized by the corresponding peak fluxes. For clear visualization, AIA lightcurves have been further normalized by 0.8, 0.85, 0.9, and 0.5 for 304 \AA , 94 \AA , 171 \AA , and 1600 \AA , respectively. Panel (b): Temporal evolution of \textit{RHESSI} and GOES X-ray fluxes in the same interval as in panel (a). \textit{RHESSI} did not observe during $\approx$01:14 UT-- $\approx$01:45 UT and $\approx$02:54 UT-- $\approx$03:25 UT. \textit{RHESSI} fluxes have been normalized by factors of $\frac{1}{2}$, $\frac{1}{20}$, $\frac{1}{10}$, $\frac{1}{100}$, and $\frac{1}{500}$ for 3--6 keV, 12--25 keV, 25--50 keV, 50--100 keV, and 100-300 keV, respectively. The red, green, and blue bars at the top of panel (b) indicate \textit{RHESSI} attenuator states A0, A1, and A3, respectively.}
\label{lightcurve}
\end{figure}

In Figure \ref{lightcurve}, we plot temporal evolution of AIA (E)UV intensities (Figure \ref{lightcurve}(a)) and \textit{RHESSI} X-ray counts (Figure \ref{lightcurve}(b)). For comparison, we have overplotted GOES SXR flux variation in both 1--8 \AA\ and 0.5--4 \AA\ channels in Figure \ref{lightcurve}(b). Comparison of GOES, \textit{RHESSI}, and AIA lightcurves clearly indicates that the pre-flare event P1 was very prominent in \textit{RHESSI} 3--6 keV and 6--12 keV channels while only the 304 \AA\ channel among other AIA channels displayed a small peak at the time of pre-flare event P1. \textit{RHESSI} did not observe during $\approx$01:13 UT-- 01:45 UT which included the pre-flare event P2. A small increment during this period was observed in the AIA 94 \AA\ and 304 \AA\ lightcurves. Interestingly, the pre-flare event P2 was more prominent in the high energy GOES channel of 0.5--4 \AA\ than the lower energy channel of 1--8 \AA . AIA lightcurves show a general agreement with \textit{RHESSI} and GOES time profiles during the main phase of the X1.0 flare. \textit{RHESSI} time profiles reveal sudden rise of HXR flux at $\approx$01:53 UT implying the onset of the impulsive phase of the X1.0 flare. The impulsive build up phase of the X1.0 flare was associated with a very interesting phenomena in the \textit{RHESSI} high energy channels of $\gtrsim$25 keV in the form of multiple short-lived spikes during $\approx$01:55 UT-- 02:00 UT (see the green, yellow and violet lines in Figure \ref{lightcurve}(b)). Here we note that, counts in the high energy channels of \textit{RHESSI} ($\gtrsim$50 keV) underwent sudden fall after $\approx$02:00 UT while counts in the low energy channels ($\lesssim$50 keV) displayed gradual decay (decay rate was faster for the 25--50 keV channel than the other channels).

\subsubsection{Pre-flare phase} \label{pre-flare}
\begin{figure}
   \centering
   \epsscale{0.85}
   \plotone{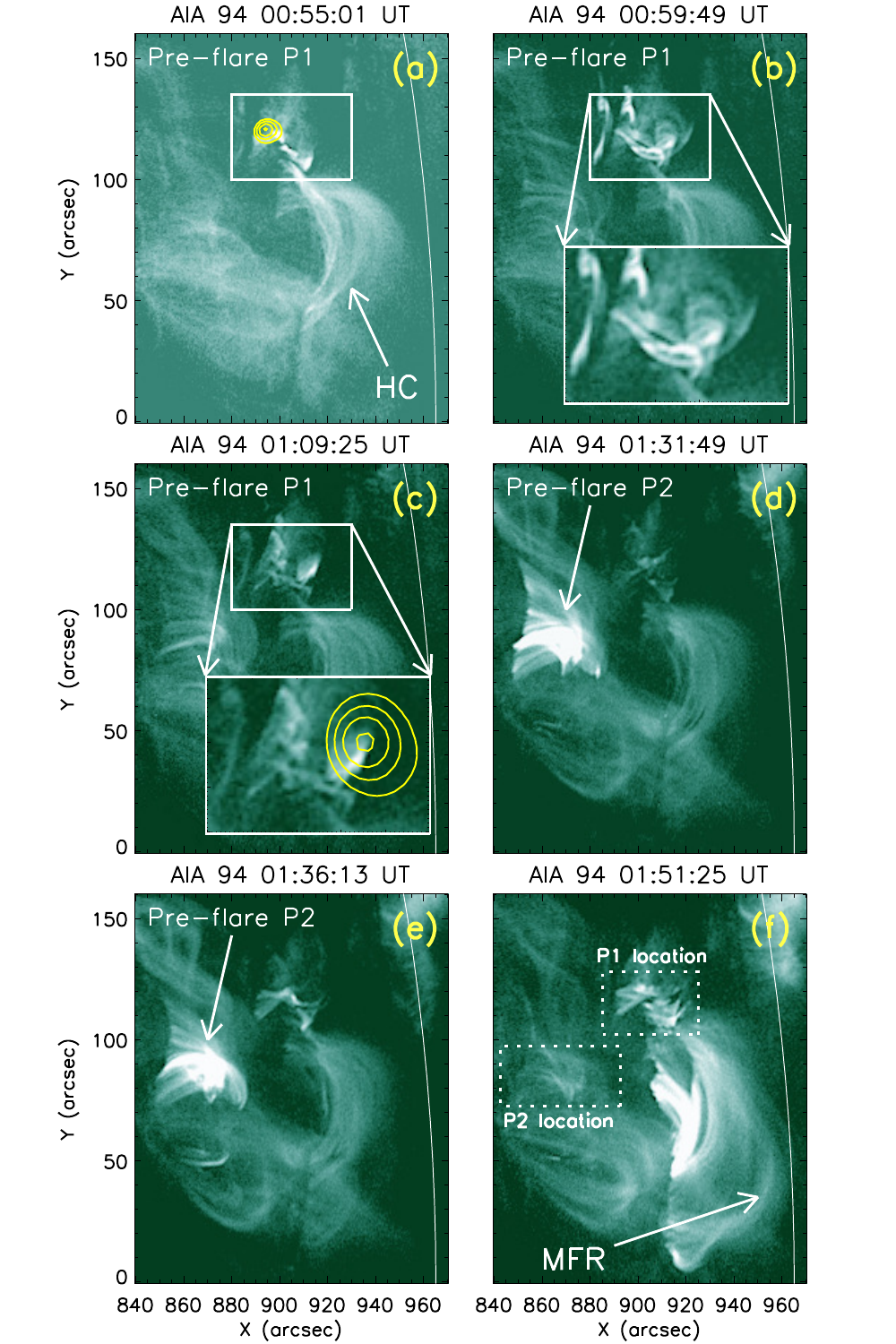}
\caption{Panels (a)--(c): AIA 94 \AA\ images showing evolution during the pre-flare event P1. During pre-flare event P1, activities were observed from a complex structure situated at the northern end of the hot channel which is outlined by the white boxes. Insets in panels (b) and (c) show zoomed in images of the region enclosed within the boxes. \textit{RHESSI} observation was available during pre-flare event P1. In panel (a) and the inset in panel (c), we plot co-temporal \textit{RHESSI} 6--12 keV contours. Contour levels are 35\%, 50\%, 70\%, and 95\% of the corresponding peak flux. Panels (d)--(e): AIA 94 \AA\ images during the pre-flare event P2. Activities during P2 were observed in the form of loop brightening which is indicated by the white arrows. Panel (f): AIA 94 \AA\ at $\approx$01:51 UT when flux rope eruption was very prominently observed in hot AIA pass band filters. The eruption of the magnetic flux rope (MFR) is indicated by the arrow in panel (f). For comparison, the locations of pre-flare activities during the P1 and P2 are shown by the white boxes with dotted lines in this panel.}
\label{pf}
\end{figure}

The two SXR pre-flare events (P1 and P2) were separated by a time span of $\approx$30 minutes (Figure \ref{goes_aia}(a) and Section \ref{S_intro}). In order to understand the location and structures of pre-flare events, we provide few representative AIA 94 \AA\ images of the activity site in Figure \ref{pf}. From these images we find that a small region situated at the north of the AR underwent a complex evolution during the pre-flare event P1 (zoomed in images of the region are shown in the insets in Figures \ref{pf}(b) and (c)). Strong \textit{RHESSI} 6--12 keV sources were observed to reside at that region during the pre-flare event P1 (co-temporal \textit{RHESSI} 6--12 keV contours are plotted in Figure \ref{pf}(a) and in the inset in Figures \ref{pf}(c)). During $\approx$01:28 UT--$\approx$01:40 UT, we observed loop brightening from a nearby eastern region within the AR which is indicated by the arrows in Figures \ref{pf}(d) and (e). Comparison of GOES time profiles (Figure \ref{goes_aia}(a)) and AIA images (Figure \ref{pf}) confirms the association of SXR pre-flare enhancements with localized energy release events within the AR. In Figure \ref{pf}(f), we plot an AIA 94 \AA\ image of the AR where the erupting hot channel can be identified clearly (indicated by the white arrow). For a comparison, we have also marked the locations of the two pre-flare events in Figure \ref{pf}(f) from which it becomes clear that the two pre-flare events prior to the X1.0 flare recorded by GOES occurred in the same AR but from two different locations. It is worth mentioning that, while the location of pre-flare event P2 displayed activity only during P2, the location of pre-flare event P1 was associated with continuous small scale activity throughout the pre-flare phase, precursor phase, and even during the impulsive phase of the X1.0 flare.

\subsubsection{Precursor phase} \label{precursor}
\begin{figure}
   \centering
   \epsscale{1.1}
   \plotone{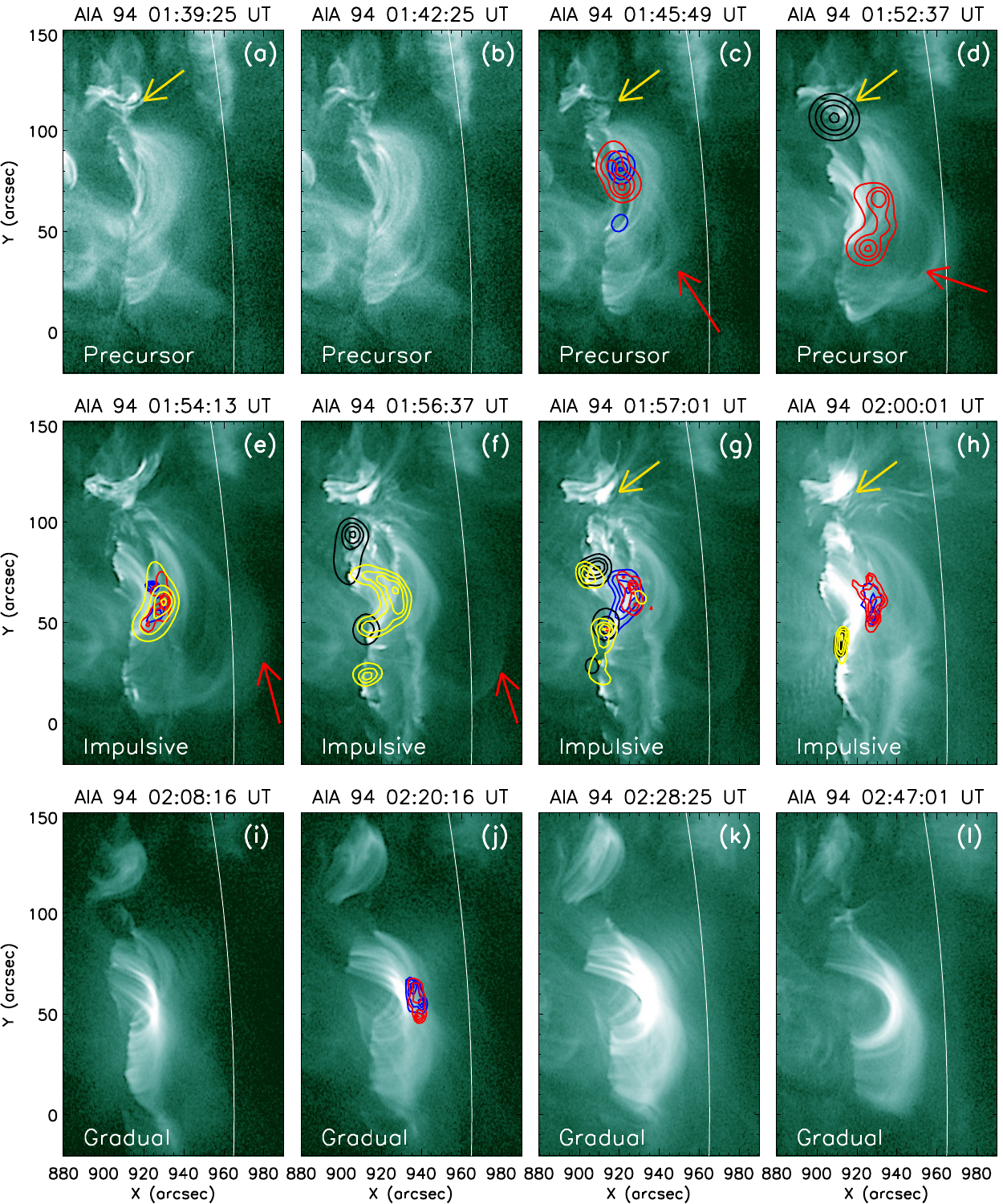}
\caption{Representative AIA 94 \AA\ images showing various evolutionary phases (precursor, impulsive, and gradual) during the X1.0 flare. The red arrows in panels (c)--(d) indicate the slow rise of the hot channel during the precursor phase which erupted during the subsequent impulsive phase of the X1.0 flare (indicated by the red arrows in panels (e) and (f)). From the location of pre-flare events `P1' (cf. Figure \ref{pf}(f)), we observed localized brightening and plasma ejection throughout the precursor and impulsive phase which is indicated by yellow arrows in panels (a), (c), (d), (g), and (h). Co-temporal \textit{RHESSI} contours of 6--12 keV (blue), 12--25 keV (red), 25--50 keV (yellow), and 50--100 keV (black) energy bands are overplotted in selected panels. The contour levels are 30\%, 50\%, 70\%, and 95\% of the corresponding peak fluxes.}
\label{f94}
\end{figure}

The X1.0 flare was associated with a distinct precursor phase during $\approx$01:37 UT to $\approx$01:53 UT when the hot channel underwent activation and exhibited a slow but steady expansion. During this phase, GOES SXR flux displayed a gradual rise in both the 1--8 \AA\ and 0.5--4 \AA\ channels (Figure \ref{goes_aia}(a)). EUV images reveal that the core field region, containing the activated hot channel, became very bright compared to the rest of the AR during this phase (Figures \ref{goes_aia}(h)--(j)). The slowly accelerating hot channel was identified in the AIA images of hot 94 \AA\ (Figures \ref{f94}(b)--(d)) and 131 \AA\ (Figures \ref{f131}(a)--(d)) channels. We found X-ray emission up to $\approx$50 keV from the hot core field region during this phase while the slowly rising hot channel steadily moved to larger heights (Figure \ref{f94}(c)--(d) and Figure \ref{f131}(b)--(d)). We observed consistent activities from the location of pre-flare event P1 that included localized brightening and plasma ejection throughout the precursor phase (indicated by the yellow arrows in Figures \ref{f94}(a), (c), and (d)). Notably, a strong 50--100 keV HXR source was observed from the location of P1 during the late precursor phase (Figure \ref{f94}(d)).

To further investigate the altitude evolution of the hot channel, we specify a narrow slit along the white line $\overline{S_1S_2}$ in Figure \ref{sli_94}(a) and plot its time evolution between $\approx$01:35 UT--02:00 UT in Figure \ref{sli_94}(b). The time-slice diagram reveals that the hot channel experienced a very gradual rise with a speed of $\approx$14 km s$^{-1}$ up to $\approx$01:53 UT. Also we note that, between $\approx$01:52 UT and 01:54 UT (the interval is marked by the two blue dotted lines in Figure \ref{sli_94}(b)), the hot channel was subjected to a rapid acceleration ($\approx$1.41 km s$^{-2}$) as the flare moved into the impulsive phase.

\subsubsection{The X1.0 flare and eruption of the hot channel} \label{fi_er}
\begin{figure}
   \centering
   \epsscale{0.9}
   \plotone{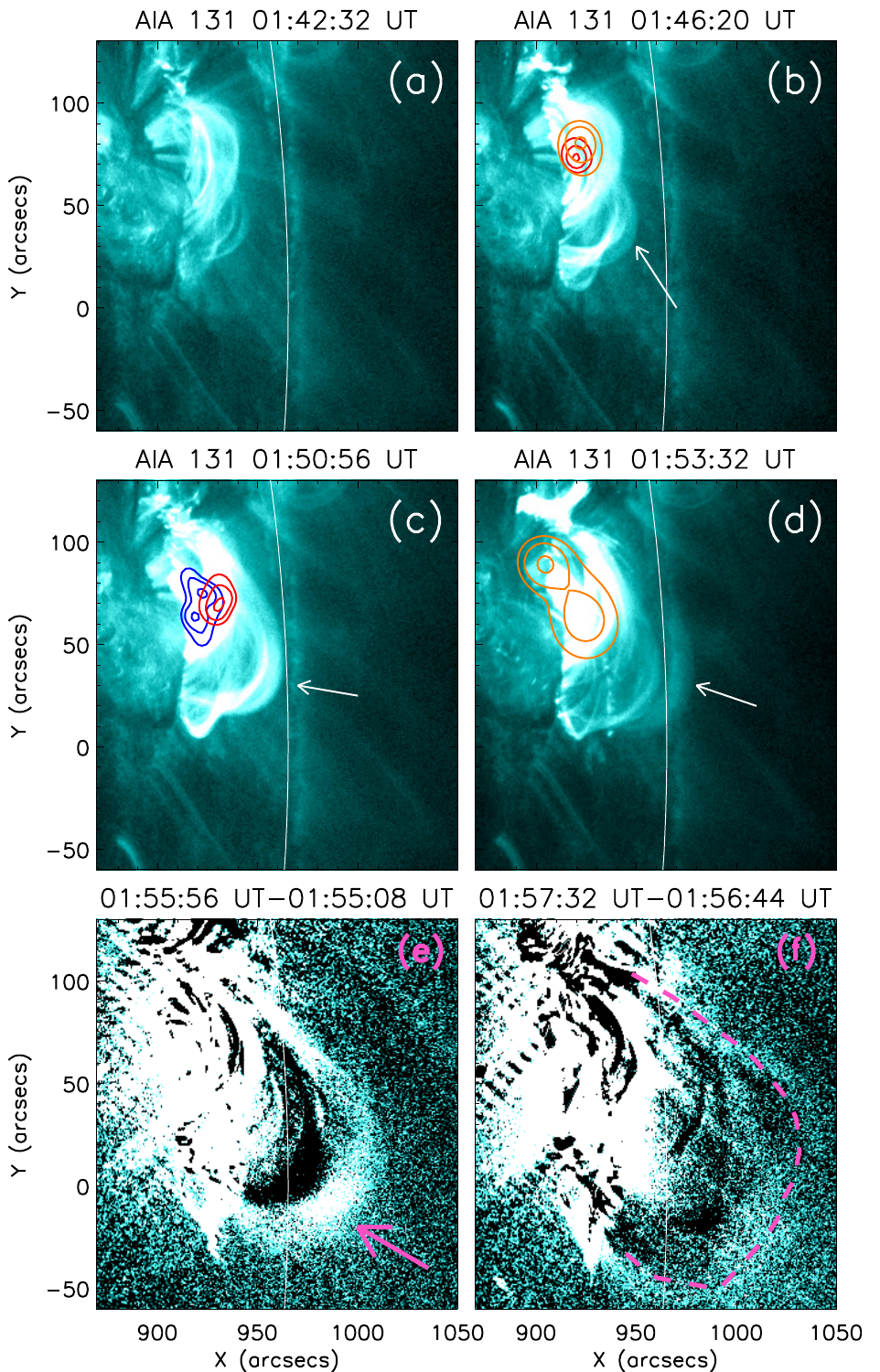}
\caption{Representative AIA 131 \AA\ direct (panels (a)--(d)) and running difference (panels (e)--(f)) images showing the hot channel eruption during the X1.0 flare. White arrows in panels (b)--(d) and pink arrow in panel (e) indicate the erupting the hot channel. In panel (f), the erupting hot channel has been outlined by a pink dashed curve. Co-temporal \textit{RHESSI} contours of 6--12 keV (blue), 12--25 keV (red), and 25--50 keV (orange) energy bands are overplotted in panels (b)--(d). Contour levels are 50\%, 70\%, and 95\% of the corresponding peak fluxes.}
\label{f131}
\end{figure}

\begin{figure}
   \centering
   \epsscale{1.1}
   \plotone{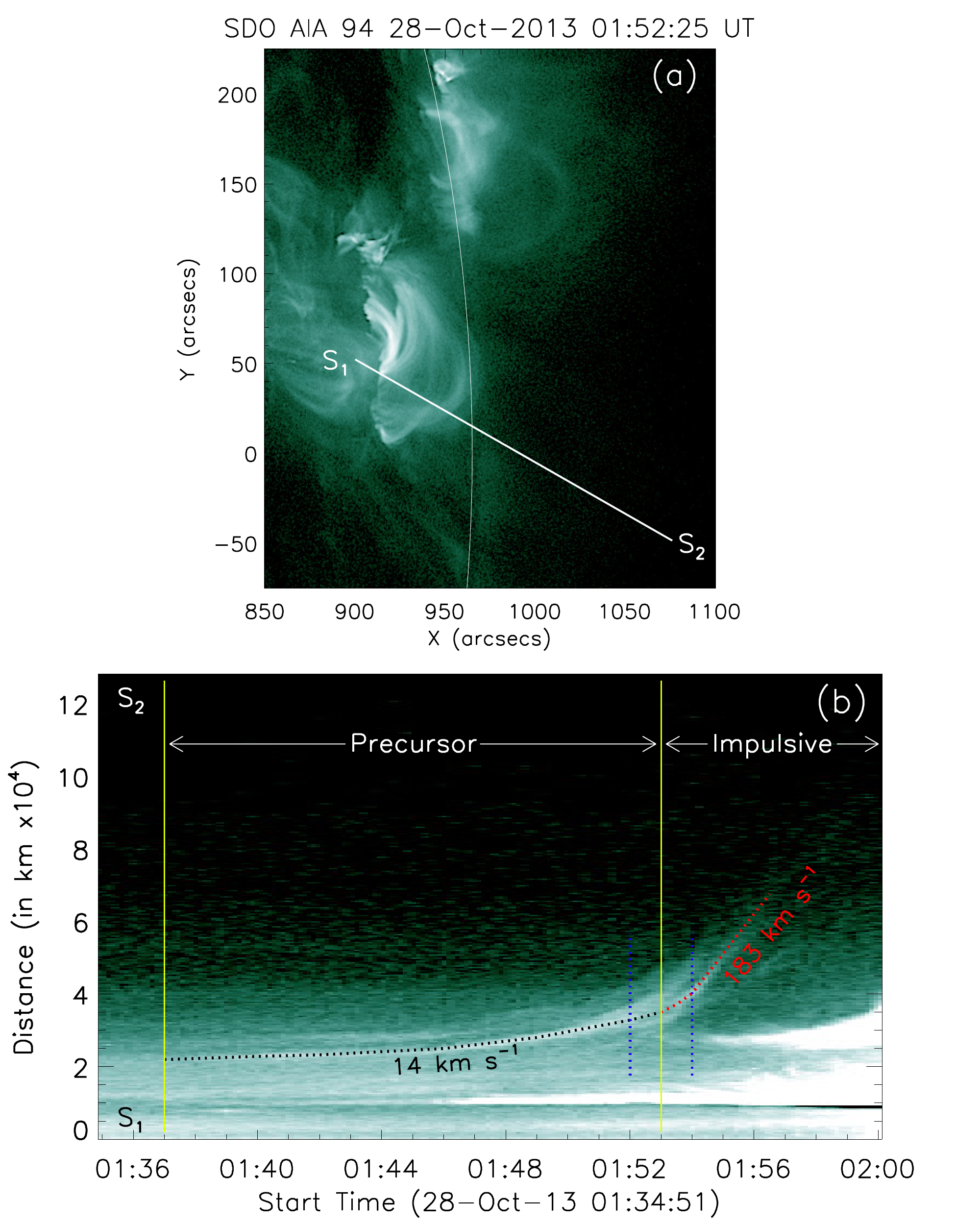}
\caption{Panel (a): An AIA 94 \AA\ image of AR11875 during the precursor phase showing a straight line along which time-slice diagram was constructed. Panel (b): Time-slice diagram showing two phase acceleration of the erupting hot channel along the slit indicated in panel (a). `S$_1$' and `S$_2$' in panels (a) and (b) indicate the orientation of the slit in the time-slice diagram. The hot channel was found to be slowly elevating during the precursor phase of the flare ($\approx$01:37 UT--01:53 UT) with a linear speed of $\approx$14 km s$^{-1}$. Afterwards, the flare entered into the impulsive phase (see Figure \ref{goes_aia}(a)) during which the hot channel underwent eruption with a linear speed of $\approx$183 km s$^{-1}$. Notably, the hot channel was subjected to a transition phase during $\approx$01:52 UT-- 01:54 UT (indicated by the vertical blue dotted lines in panel (b)) when it moved continuously from the state of slow rise to eruptive expansion with a rapid acceleration of $\approx$1.41 km s$^{-2}$.}
\label{sli_94}
\end{figure}

The impulsive phase of the X1.0 flare started at $\approx$01:53 UT (Figure \ref{goes_aia}(a)). During this interval, the slowly rising hot channel entered into the eruptive phase (Figure \ref{sli_94}(b)). As the hot channel started erupting, formation of post-flare arcade began underneath it. The co-temporal \textit{RHESSI} images reveal intense HXR emission up to $\approx$50 keV energies from the apex of the post-flare arcade (Figures \ref{f94}(e)--(h)). During the peak phase of the X1.0 flare, along with the coronal HXR sources, strong HXR sources up to $\approx$100 keV energies were observed from the footpoint of the arcades (Figure \ref{f94}(f)--(h)). As the hot channel moved into higher coronal heights, it lost its brightness and slowly became indistinct in direct images after $\approx$01:56 UT (the erupting hot channel is indicated by red arrows in Figure \ref{f94}(e)--(f). In Figures \ref{f131}(e)--(f), we plot running difference AIA 131 \AA\ images where we indicate the hot channel by a pink arrow and a dashed pink line, respectively. The flare moved into the gradual phase after $\approx$02:02 UT. During this phase, the post-flare arcade slowly increased in height and continued to emit hot, intense diffused emission (Figures \ref{f94}(i)--(l)). \textit{RHESSI} sources of energies up to 25 keV were observed from the top of the post-flare arcade (Figure \ref{f94}(j)).

\begin{figure}
   \centering
   \epsscale{1.1}
   \plotone{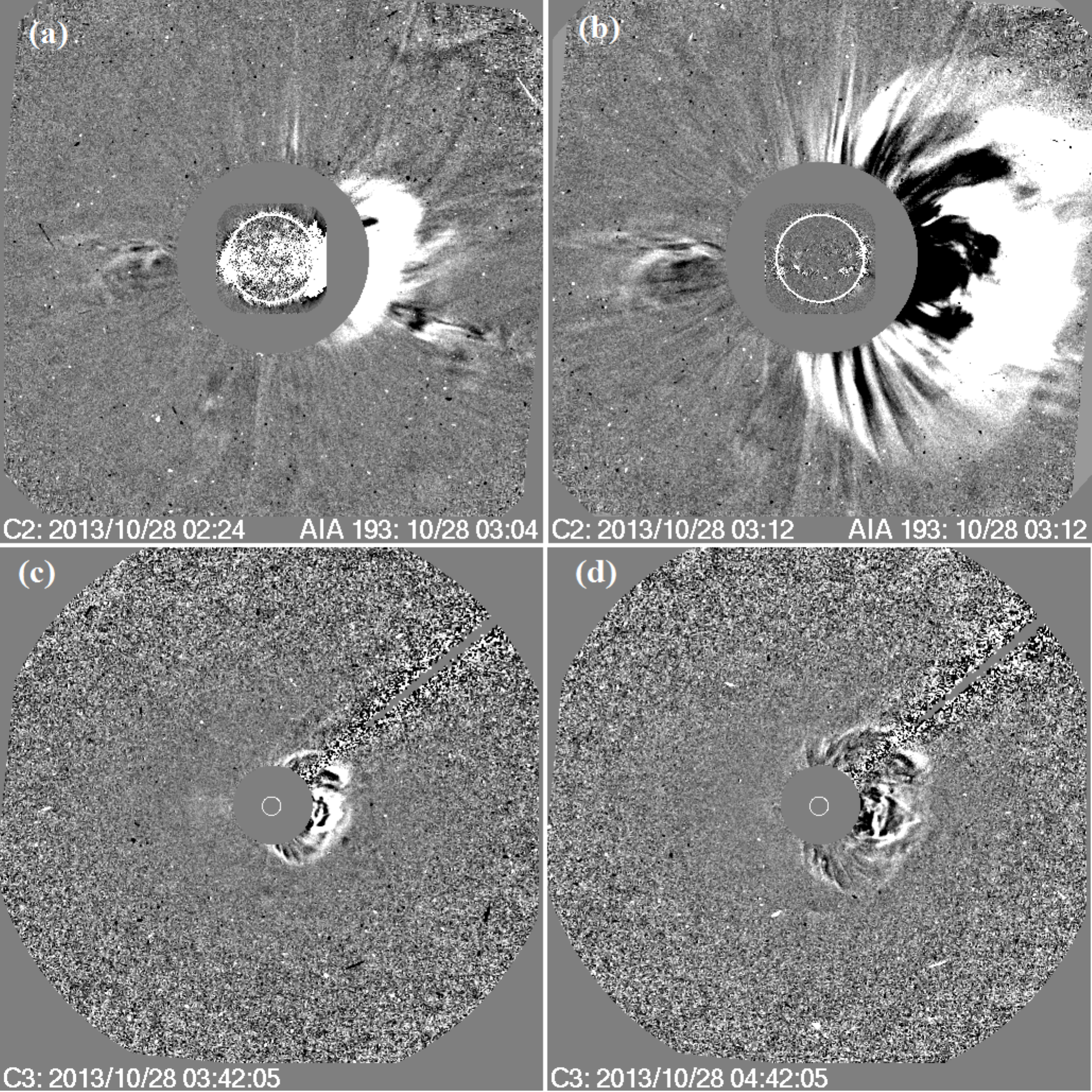}
\caption{Running difference images obtained from LASCO C2 (panels (a), (b)) and C3 (panels (c), (d)) showing the propagation of the halo CME originated during the X1.0 flare from AR11875. The linear speed of the CME calculated within the LASCO field of view is 695 km s$^{-1}$.}
\label{cme}
\end{figure}

The time-slice diagram (Figure \ref{sli_94}(b)) suggests that after undergoing a slow expansion during the precursor phase, the hot channel exhibited fast eruption with a linear speed of $\approx$183 km s$^{-1}$ during the impulsive phase. The eruption of the hot channel resulted in a halo CME which was captured in the LASCO C2 and C3 coronagraph observations (Figure \ref{cme}). The CME was first detected by LASCO C2 at 02:24 UT\footnote{\url{https://cdaw.gsfc.nasa.gov/CME_list/UNIVERSAL/2013_10/yht/20131028.022405.w360h.v0695.p296g.yht}} when the leading edge of the CME reached a coronal height of $\approx$3.82 R$_\odot$. LASCO C3 observed the CME until 07:30 UT at $\approx$22.91 R$_\odot$. Within the field of view (FOV) of LASCO, the CME was propagating toward the position angle 296$^\circ$ with a linear speed of $\approx$695 km s$^{-1}$ and a slow deceleration of 12.1 m s$^{-2}$.

\subsection{\textit{RHESSI} spectroscopy} \label{rhessi}
To study the HXR spectral evolution of the X1.0 flare during the impulsive phase, we conducted \textit{RHESSI} spectroscopy. For the purpose, we used energy bins of $\frac{1}{3}$ keV from 6 to 15 keV, 1 keV from 15 to 100 keV, and 5 keV from 100 to 300 keV energies. At the start of the impulsive phase of the X1.0 flare, the attenuator of \textit{RHESSI} was set at A1 (i.e., thin shutter; Figure \ref{lightcurve}(b)). At $\approx$01:56 UT, the attenuator state was changes to A3 (i.e., both thin and thick shutters) and remained so until $\approx$02:09 UT except two short intervals of $\sim$1 minute (at $\approx$02:00 UT and $\approx$02:05 UT) when the attenuator state was changed to A1. After $\approx$02:09 UT, the attenuator state was changed to A1 as the flare had moved into the gradual phase. The time intervals for spectral analysis were chosen to be 20 s. Notably, \textit{RHESSI} measurements during the impulsive phase of the X1.0 flare were contaminated by two periods of low energy radiation-belt particle events. To restrict the spectroscopic measurements from contamination due to the particle events, we confined our spectroscopic analysis  between 01:57 and 02:01 UT. Spectral fits were obtained using a forward-fitting method implemented in the idl-based Object Spectral Executive code \citep[OSPEX;][]{Schwartz2002}. For creating the spectrum, we used the combined \textit{RHESSI} front detectors 1--9 excluding 2 and 7 \citep[see][]{Smith2002, Holman2011}. Two fitting models were used: line emission from an isothermal plasma and thick-target bremsstrahlung from non-thermal electrons interacting with the chromosphere \citep{Holman2003}. From these fits, we derived the temperature (T) and emission measure (EM) of the hot flaring plasma, as well as the non-thermal electron spectral index ($\delta$) for the non-thermal component and break energy (E$_\text{B}$) between the thermal and the non-thermal components.

\begin{figure}
   \centering
   \epsscale{1.1}
   \plotone{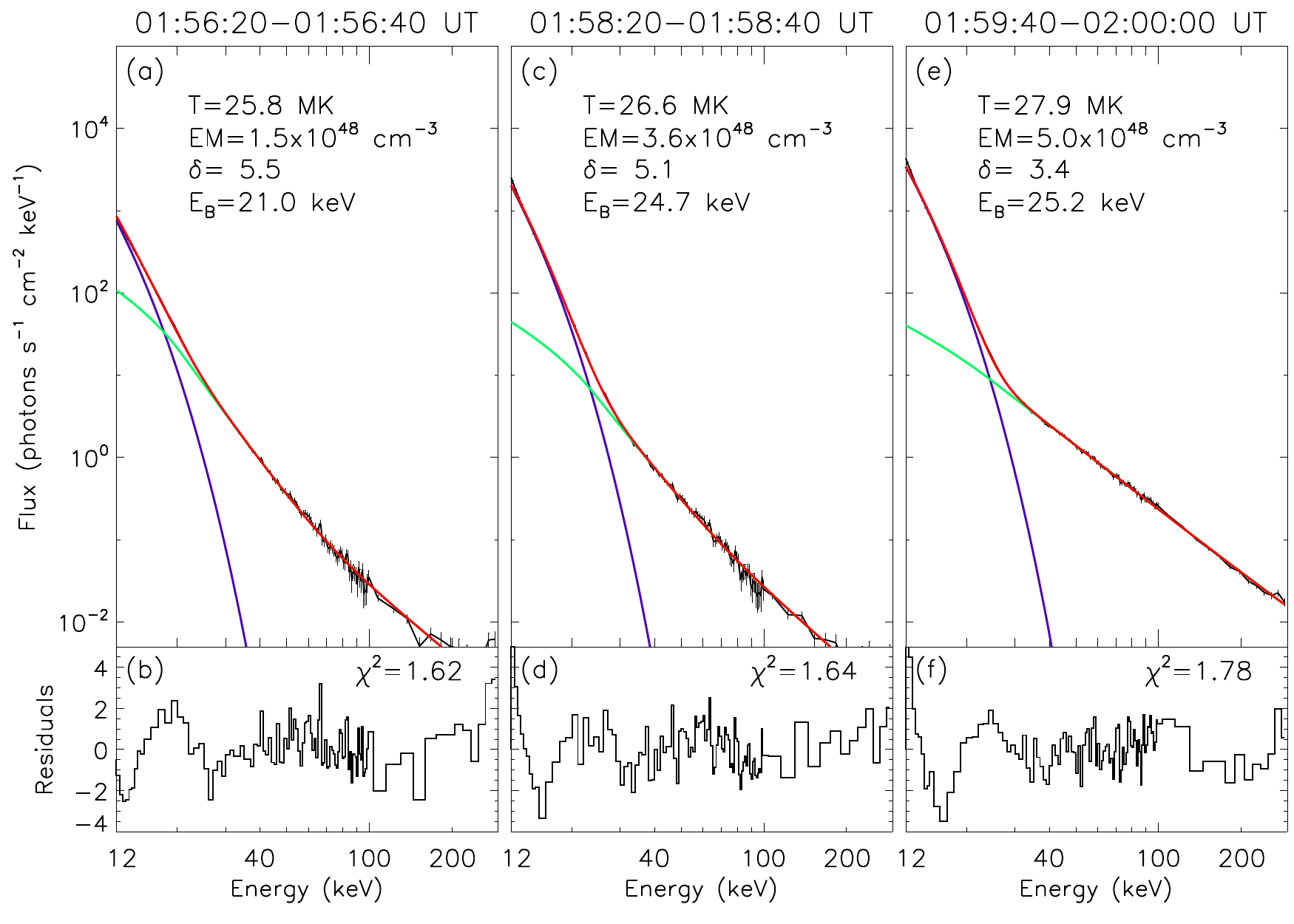}
\caption{Representative \textit{RHESSI} X-ray spectra during the impulsive phase of the X1.0 flare along with their respective residuals. These spectra were fitted with a combination of an isothermal component (blue line) and a thick-target bremsstrahlung model (green line). The red lines indicate the sum of the two components. The energy range chosen for fitting is [12--295] keV. The notations used in different panels are as follows: T: temperature obtained from the isothermal component, EM: emission measure, $\delta$: power law index for the non-thermal fitting, $\text{E}_\text{B}$: break energy separating thermal energy from non-thermal energy, $\chi^2$: goodness of the fitting.}
\label{spec}
\end{figure}

In Figure \ref{spec}, we show few spatially integrated, background-subtracted \textit{RHESSI} spectra along with their respective fits and residuals for three selected intervals. We find that both temperature and emission measure increased during the impulsive phase and reached at $\approx$28 MK and $\approx$5$\times$10$^{48}$ cm$^{-3}$, respectively, at the peak phase. At this time, as expected, with the evolution of the flare during the impulsive phase, the low energy cut-off (i.e., break energy E$_\text{B}$) increased progressively up to $\approx$25 keV. During the impulsive rise and peak of hard X-ray flux (Figures \ref{spec}(c) and (e)), the hardness of non-thermal component increased highly with spectral index ($\delta$) $\approx$3.4.

\subsection{Dynamic radio spectrum} \label{radio}
\begin{figure}
   \centering
   \epsscale{1.1}
   \plotone{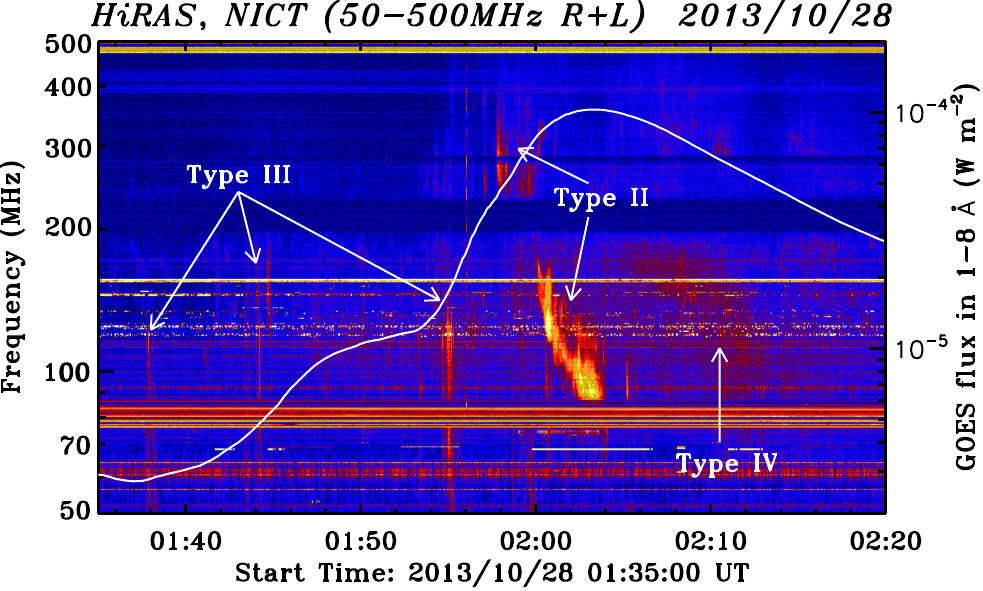}
\caption{Dynamic radio spectrum recorded by the HiRAS spectrograph on 2013 October 28 from 01:35 UT-- 02:20 UT within the frequency range 50--500 MHz, showing many discrete type III bursts between $\approx$01:37 UT--01:55 UT, a split-band harmonic of type II burst between $\approx$01:58 UT--02:04 UT and a faint type IV spectra during $\approx$02:05 UT--02:14 UT. For reference, we have overplotted GOES SXR flux variation in the 1--8 \AA\ range by the white curve.}
\label{hiras}
\end{figure}

The impulsive phase of the X1.0 flare exhibited a spectacular display of different types of radio structures in the dynamic spectra (Figure \ref{hiras}). A series of type III bursts were recorded by HiRAS during $\approx$01:37 UT--01:55 UT followed by a split-band type II burst. Since, the AR was situated at the limb, it is likely that the observed split-band of type II was the harmonic as the fundamental part of type II bursts originated near solar limbs get severely attenuated \citep{Gopal2013}. A brief brightening was recorded at $\approx$01:58 UT within frequency range $\approx$210--400 MHz. Observation within the frequency range $\approx$200--210 MHz was rejected throughout the interval by HiRAS. The type II burst was very clear between $\approx$02:00 UT--02:04 UT within the frequency range $\approx$75--180 MHz. However, from the apparent drift of the type II bands and dynamics of the erupting hot channel, it can be understood that the brief brightening within frequency range $\approx$210--400 MHz was an early extension of the type II. The type II burst was followed by a faint dynamic type IV signatures between $\approx$02:05 UT and 02:14 UT within the frequency range $\approx$70--180 MHz.

\section{Discussion} \label{disc}
In this article, we present a multi-wavelength analysis of the evolutionary phases of an X1.0 flare associated with the activation and eruption of a hot channel from active region NOAA 11875. On the day of reported activity i.e., 2013 October 28, the AR was situated near the western limb of the Sun at heliographic co-ordinates of $\sim$N07W66. We provide a detailed investigation of the activities during the pre-flare phase, a relatively less studied and ill-understood aspect of the overall flare evolution. The activation and steady, slow rise of the pre-existing hot channel accompanied the precursor phase of the X-class flare while its eruption proceeded with the impulsive and peak phases. We have summarized different evolutionary phases of the X1.0 flare in Table \ref{table}. 

Our study distinctively reveals the manifestation of a prolonged pre-flare phase of $\approx$52 min duration ($\approx$00:45 UT-- $\approx$01:37 UT; Figure \ref{goes_aia}). Further, the pre-flare phase is characterized by two distinct SXR peaks, P1 and P2, occurred $\approx$38 min and $\approx$7 min, respectively, prior to the onset of the flux rope activation. Comparison of GOES time profiles with AIA EUV images confirmed that both the pre-flare events occurred from different locations within the AR (Figure \ref{goes_aia}). While the location of the first pre-flare event (i.e., P1) was adjacent to the hot channel, location of the second pre-flare event (i.e., P2) was a distant one. Notably, while the location of P2 was associated with EUV brightening during the pre-flare event P2 only, the location of P1 experienced continuous EUV brightening throughout the pre-flare phase and even during the later precursor and impulsive phases of the flare. Statistical studies, conducted with the \textit{Yohkoh}-SXT images, of the pre-flare locations with respect to the main flare sites by \citet{Farnik1996, Farnik1998} revealed abundance of `adjacent' cases while the `distant' pre-flare sources were found to be rather uncommon. However, while discerning the locations of pre-flare localized emissions over the large-scale main flare brightenings, we should also consider that the \textit{SDO}/AIA images have much better temporal and spatial resolutions compared to the \textit{Yohkoh} images.

Pre-flare activities prior to a flare are considered to be important in order to understand the physical conditions that lead to flares and associated eruptions \citep{Farnik1996, Farnik1998, Farnik2003, Chifor2006, Chifor2007, Joshi2011, Joshi2013}. Contemporary studies, dealing with the analysis of X-ray images and spectra during the mild pre-flare X-ray emission \citep{Joshi2013, Hernandez2019}, suggest pre-flare phase to be associated with the events of small-scale magnetic reconnection that can potentially destabilize the flux ropes as described in different models of solar eruptions (i.e., tether-cutting and breakout reconnection models). Our analysis suggests that the location of the pre-flare event `P1' was associated with consistent EUV brightening along with thermal X-ray sources (Figure \ref{pf}) suggesting continuous small-scale reconnection events in a localized region that resulted into the eruption of the hot channel (i.e. the pre-existing flux rope). Our results can be interpreted in the light of multiple pre-flare events prior to a major flare of X-class analyzed by \citet{Joshi2013}. Their work revealed that the activities during the pre-eruption phase were characterized by three localized episodes of energy release occurring in the vicinity of a filament that produced intense heating along with non-thermal emission. Based on these observations, \citet{Joshi2013} concluded that localized magnetic reconnections beneath the flux rope in the pre-eruption phase played an important role in destabilizing the active region filament through the tether-cutting process. From the spatial and temporal characteristics of pre-flare activities occurring at the location P1, we conclude that continuous, small-scale reconnection events in the vicinity of the northern leg of the hot channel would result in restructuring of the magnetic field configuration in a localized region besides reducing the magnetic flux near one of the footpoints of the magnetic flux rope (MFR). This process would progressively lead toward establishing the conditions favorable for the destabilization of the MFR and subsequent large-scale eruption. 

An important aspect of present observations lies in the detection of a stable, hot EUV channel that pre-existed in the active region corona at least $\approx$67 min before its activation during the precursor phase (Figure \ref{I_intro}(d)). Contemporary studies have established that hot coronal channels are one of the observational evidences of MFRs \citep[see e.g.,][]{Chen2011, Cheng2014, Songb2015, Joshi2017, Joshi2018, Mitra2018}. Notably, observations showing prior formation of flux ropes that remain stable during the pre-eruption phase are still uncommon with availability of only a few such reported incidences \citep{Cheng2014, Song2015}. The exact topology and formation process of MFRs are still unclear and debatable. According to the mechanism proposed by \citet{Van1989}, flux cancellation through the photospheric converging and shearing motions leads into the formation of MFR which has been subsequently supported by the numerical studies by \citet{Amari1999, Amari2000} where the twisted flux rope remains stable for a period before erupting. Many studies have also recommended the in-situ formation of MFRs in the corona by magnetic reconnection, i.e., the MFR builds up with the initiation of the flare itself \citep{Cheng2011, Song2014b, Chintzoglou2015, Wang2017}. Another class of model suggests that MFRs are generated in the convection zone of the Sun and then emerges into the solar atmosphere by buoyancy \citep{Caligari1995, Fan2001, Archontis2004, Martinez2008}. With the detection of a pre-existing hot channel right from the beginning of the studied interval and subsequent occurrence of small-scale energy release processes at adjacent and remote locations, our observational results support the scenario of flux rope emergence from the convection zone. During the impulsive phase of the X1.0 flare, the hot channel became activated and erupted resulting into a halo CME (Figures \ref{f94}, \ref{f131}, and \ref{cme}).  With high proneness towards eruption and association with CMEs, hot channels are believed to be the earliest signatures of CMEs \citep{Canfield1999, Pevstov2002, Gibson2004, Gibson2006b, LiuC2007, Liu2010, Cheng2014b, Kumar2016, Joshi2017, Joshi2018, Liu2018, Mitra2018, Hernandez2019}; our observations are consistent with these earlier studies.

We would like to further emphasize on the early dynamical evolution of the hot channel which is characterized by its slow yet steady rise with a speed of $\approx$14 km s$^{-1}$. This phase lasted for a period of 16 minutes during which the intensity of the flaring region increased gradually. The temporal association between the slow rise phase of the MFR and the precursor flare emission is of great significance toward understanding the origin of CMEs \citep{Zhang2001, Zhang2004, Zhang2006}. The early slow rise phase essentially marks the CME initiation in the source active region which is found to be well correlated with the mild and consistent precursor phase emission, commonly observed as gradually rising SXR flux \citep{Zhang2006}. It has been suggested that the precursor phase activities correspond to the dynamical formation of a current sheet underneath the MFR that subsequently reconnects to trigger the onset of the main phase of the eruptive flare \citep{Zhou2016}. Our analysis readily supports this scenario as the hot channel underwent an abrupt transition from the state of slow rise to the fast acceleration as a continuous process which precisely bifurcates the precursor and impulsive phase of the eruptive flare (cf. Figures \ref{sli_94}(b) and \ref{goes_aia}(a)).

The early impulsive phase and the eruption of the flux rope was associated with strong HXR sources of energies up to $\approx$100 keV (Figure \ref{f94}). We further emphasize on the appearance of a very high energy HXR source at the adjacent pre-flare location (Figures \ref{f94}(d), \ref{pf}). Appearance of such high energy HXR sources from a pre-flare location is rather surprising. In addition, we note that HXR sources were primarily concentrated at the northern part of the AR which is also the pre-flare location P1 (Figure \ref{pf}(a), (c) and \ref{f94}(c)--(d)) suggesting strong reconnection and restructuring of the magnetic field lines. This is supported by small-scale plasma ejection from this region (Figure \ref{f94}). 

During the late impulsive phase which is accompanied with fast eruption of the flux rope, non-thermal high energy HXR sources ($>$50 keV) were observed from the footpoint location while lower energy X-ray sources ($\lesssim25$ keV) were observed from the apex of the post flare arcade (Figure \ref{f94}). Notably, 25--50 keV HXR sources, probably containing mix of thermal and non-thermal emissions, were found from both looptop and footpoint locations (Figures \ref{f94}(e)--(h)). High energy HXR footpoint sources can be interpreted in terms of `thick target bremsstrahlung' model \citep{Brown1971, Hudson1972, Syrovatskii1972, Emslie1983}. According to this model, highly accelerated electrons from the site of magnetic reconnection in the corona are injected downward into the transition region and the chromospheric layers along the post reconnection magnetic field lines with relativistic speed. The relativistic electrons collide with cool and dense chromospheric plasma and deposit their energy with the production of HXR footpoint sources \citep{Korchak1967b, Korchak1967a, Cline1968, Moza1986, Brown2002, Flannagain2015, Reep2016}. Two basic models of HXR flares-- impulsive injection and continuous injection models differ on the time-scale of electron acceleration. According to the impulsive injection model proposed by \citet{Tatakura1966}, electrons are accelerated in multiple episodes of short time-scales in magnetic traps (i.e., magnetic loops). In contrast, the continuous injection model suggests continuous acceleration of electrons in dense chromosphere \citep{Kundu1963, Action1968, Kane1970, Brown1971}. In the view of multiple spikes in HXR time profiles during the impulsive phase (Figure \ref{lightcurve}(b)) and multiple type III bursts observed during the build up phase of the X1.0 flare (Figure \ref{hiras}), our analysis suggests occurrence of discrete events of particle acceleration during the impulsive phase. The overall distribution of low energy looptop and high energy footpoint X-ray sources, observed in our work, is in agreement with the CSHKP model of solar eruptive flares and also consistent with the general energy release scenario developed during the \textit{RHESSI} era \citep{Krucker2003, Veronig2006, Joshi2007, Joshi2013}. 

During the impulsive phase of the X1.0 flare, the MFR underwent fast acceleration ($\approx$1.41 km s$^{-2}$) with the enhancement of its eruption speed to $\approx$183 km s$^{-1}$. The brief but rapid acceleration phase of the flux rope is co-temporal with the onset time of multiple HXR bursts observed in the high energy channels ($\gtrsim$25 keV) of \textit{RHESSI} (cf. Figures \ref{sli_94}, \ref{f94} and \ref{lightcurve}(b)) and a series of type III radio bursts (Figure \ref{hiras}). Type III radio bursts are produced by near-relativistic electrons propagating along the open magnetic field lines in the corona, restructured by the magnetic reconnection \citep{Bastian1998, Reid2014, ChenB2018}. The appearance of strong looptop and footpoint HXR sources during the impulsive phase implies rapid dissipation of magnetic energy in the current sheet as a result of an increase in the rate of magnetic reconnection \citep{Sui2003} which is also likely to be responsible for the fast acceleration of the flux rope. While studying full CME kinematics including the initiation and impulsive acceleration phase of two fast halo CMEs and the associated flares, \citet{Temmer2008} found a close synchronization between the CME acceleration profile and the flare energy release as indicated by the \textit{RHESSI} HXR flux onsets. They interpreted their results in terms of a feedback relationship between CME dynamics and reconnection events in the current sheet beneath the CME. Similar results were reported by many subsequent studies \citep[see e.g.,][]{Zhang2012, Songa2015, Joshi2016}.

\section{Summary and conclusions} \label{summary}
We provide a comprehensive investigation of the unveiling and subsequent eruption of an MFR which we observationally recognized in the form of an EUV hot channel. This quasi-static hot channel was observed from at least 67 min before its activation, evidencing the pre-existence of the MFR in the corona much prior to the CME initiation.  These observations readily support the idea that hot channel structures can be regarded as the earliest signatures of a CME in the source active region. Our work also focuses on the pre-flare activity and its role in the destabilization of a stable MFR. The prolonged pre-flare phase exhibited two distinct SXR peaks which are characterized by small-scale energy release processes within the confined regions. With respect to the pre-existing MFR, one of the pre-flare events occurred at an `adjacent' location while the second one was a `remote' activity but within the active region. Importantly, the adjacent pre-flare event took place near one of the footpoints of the MFR. The activation of MFR occurred with the strong intensity enhancement at the adjacent pre-flare activity location at HXR energies besides EUV brightening. These observations imply toward the role of pre-flare magnetic reconnections in the destabilization of a stable MFR. The destabilization of an MFR can be characterized by a gradual rise in the GOES SXR flux during the precursor phase of a flare accompanied by slow elevation of the MFR. A current sheet is formed underneath the destabilized MFR that subsequently reconnects to trigger the onset of the `standard' flare. With the signatures of the initiation of large-scale magnetic reconnection triggered by the eruptive MFR, such as, high energy non-thermal HXR sources and multiple type III radio bursts, the MFR underwent a transition from slow to fast motions which points toward a feedback relationship between the initial CME acceleration and strength of the large-scale magnetic reconnection. 

\acknowledgments The authors would like to thank the \textit{SDO} and \textit{RHESSI} teams for their open data policy. \textit{SDO} and \textit{RHESSI} are NASA's missions under the Living With a Star (LWS) and SMall EXplorer (SMEX) programs, respectively. The authors would also like to thank the NICT team for providing the HiRAS dynamic spectrum. The authors acknowledge the constructive comments and suggestions of the anonymous referee, which improved the presentation and scientific content of the article.

\end{document}